\documentstyle [12pt]{article}
\begin{document}
\input epsf

\hfill {WM-01-108}

\hfill {\today}

\vskip 1in   \baselineskip 24pt

{
\Large
   \bigskip
   \centerline{Large Electric Dipole Moments of Heavy Leptons}
 }

\vskip .8in
\def\bar{\overline}

\centerline{Marc
Sher\footnote{Email: sher@physics.wm.edu}  }
\bigskip
\centerline {\it Nuclear and Particle Theory Group}
\centerline {\it Physics Department}
\centerline {\it College of William and Mary, Williamsburg, VA 23187, USA}

\vskip 1in

{\narrower\narrower In many models of CP violation, the electric dipole moments (EDMs) of 
leptons scale as the cube of the lepton mass.  In these models, the EDM of a $100$ GeV heavy
lepton would be a billion times greater than that of the muon, and could be as large as
a $0.01$ e-fermi.   In other models, in which the heavy leptons have different properties from the 
lighter generations,  a similarly large EDM can be obtained.  A large EDM could dominate the electromagnetic
properties of heavy leptons.   The angular distribution and production cross-section of both charged and neutral
heavy leptons with large dipole moments is calculated and discussed.  The interesting possibility
that a heavy neutrino with a large EDM could leave an ionization track in a drift chamber is
investigated. }
\newpage
\section{Introdution}

The origin of the fermion masses and mixing angles remains a mystery.  
Many models for explaining their values exist, most involving 
additional symmetries, but it will require additional experimental data to 
distinguish between them.  Much of the excitement about the
discovery of neutrino masses and mixing angles is due to the hope
that their values will provide clues toward solving the mystery of flavor.

Another potential source of information may come from electric and magnetic 
dipole moments.  These moments may also reflect the flavor structure of the
model.  Just as the Yukawa couplings form a matrix in generation-space, the 
interaction of two fermions with a photon will also be a matrix in 
generation-space.  The real and imaginary parts of the diagonal elements will 
lead to the magnetic and electric dipole moments; the off-diagonal elements
will lead to decays such as $\mu\rightarrow e\gamma$.   Measurements of these
moments will give valuable clues towards understanding flavor physics, and
will give hints of physics beyond the standard model (as demonstrated by the
great excitement over the apparent discrepancy\cite{e821} with the muon's magnetic
dipole moment).

In this paper, I will focus on the electric dipole moments (EDM's) of leptons.
Here, there is the possibility of substantial improvement in the experimental
bounds (or discovery) in the near future.  A recently approved experiment\cite{neweedm}
expects to lower the current bound\cite{commins} on the electron EDM of $4.3 \times
10^{-27}$  e-cm to at least $10^{-29}$ e-cm and possibly a couple of orders of magnitude
better.  The muon EDM limit\cite{bailey} is now $1.1\times 10^{-18}$ e-cm, but proposals
to lower this by six orders of magnitude exist\cite{newmuedm}.  The tau limit is now 
listed\cite{pdg} at $3\times 10^{-16}$ e-cm, but combining limits on the weak dipole
moment with $U(1)$ symmetry improves that by a factor of thirty\cite{masso}.

What values might one expect for the EDM's?  In the Standard Model, the
EDM's are extremely small\cite{smedm}.  However, in most extensions of the Standard Model,
they are substantially larger.    In multi-Higgs
models, the EDM of the muon can be as large\cite{barger} as $10^{-24}$ e-cm, and thus within
reach of currently planned experiments.  In leptoquark models, the muon and tau EDMs are
typically $10^{-24}$ e-cm and $10^{-19}$ e-cm, respectively\cite{bern}.  In left-right
models\cite{lrmodels}, the muon EDM is  typically $10^{-24}\sin\alpha$ e-cm, where $\alpha$
is a phase angle.  In the MSSM\cite{mssm}, the electron EDM is somewhat above the
experimental bounds if the phases are all of order unity.  Thus, we see that a wide variety
of models give EDMs that can be observed in the next round of experiments.

Recently, Babu, Barr and Dorsner\cite{bbd} discussed how the EDM's of leptons scale
with the lepton masses.  In many models, such as the MSSM, they scale linearly
with the mass.  However, in a number of models, such as some multi-Higgs,
leptoquark and  flavor symmetry models, the EDM scales as the cube
of the lepton mass.  In these models the tau EDM will be $5,000$ times larger
than the muon EDM.  (It should be noted that the electron EDM in some of these models
receives a two-loop contribution which only varies linearly, and thus it
need not be negligible.)

The purpose of this paper is to point out that if a fourth generation lepton 
doublet exists, then reasonable models exist in which  one would
expect an enormous EDM.  For example, in models with cubic scaling, a muon EDM in the expected
range, $d_\mu = 10^{-24}$ e-cm, would mean that  a $100$ GeV lepton would have an EDM 
of  $0.01$ e-fermi.  Such an EDM will dominate the
electromagnetic interactions, dramatically changing the phenomenology of
such leptons.   Even more interesting is the possibility that the heavy
neutrino (also expected in the $100$ GeV range) could have an enormous EDM,
leading to the question of whether such a neutrino could leave an ionization
track.

How realistic is the possibility of such a large EDM for a heavy lepton?  A 
specific model with a cubic scaling is the model of Bernreuther, Schroder and 
Pham\cite{bsp}, in which CP violation occurs in the Higgs sector.  In this 
model, the EDM of a heavy lepton (they considered the top quark, but the 
results are similar) is constrained by the electron EDM and loop diagrams, and
the maximum EDM for a heavy lepton is a factor of 100 below $0.01$ e-fermi.  
In other models, such as the model of Ref. \cite{bbd} in which one has their parameter $c=0$, the electron EDM does not pose such a constraint and a large EDM is allowed.  If we do not rely on any specific model, then if one simply
assumes that CP violation is due to new physics at a TeV scale, and writes
an effective dimension-five Lagrangian as ${c\over \Lambda}\bar{L}_L\sigma^{\mu\nu}i\gamma_{ 5}L_RF_{\mu\nu}$, then the added assumption that all particles
with $O(100)$ GeV masses have couplings of $O(1)$ (which is true for the $W$, $Z$ and top quark) results is a very large EDM of the order of $0.01$ e-fermi.

As an existence proof, one could assume that the fourth generation leptons form
a vectorlike isodoublet.  This could explain the large neutrino mass.  If they
couple to singlet Higgs fields with large, complex VEVs, then a large EDM
could easily be generated.  Such a model is not a cubic scaling model--the
EDMs of the light fermions remain negligible.
Our main point is that such a large EDM is certainly not excluded, and we thus
investigate its consequences.

\section{Heavy Lepton Production}

The most dramatic effect of a large EDM of a heavy lepton will be in the 
production cross-section and angular distribution.  As pointed out by Escribano
and Masso\cite{masso}, the $U(1)$ invariant effective operator giving an EDM is
given by $\bar{L}_L\sigma^{\mu\nu}i\gamma_{ 5}L_RB_{\mu\nu}$, where $B_{\mu\nu}$ 
is the $U(1)$ field tensor.  This will lead to a coupling to the photon, 
which we define to be the EDM, as well as a coupling to the $Z$ which is given
by the EDM times $\tan\theta_W$.  We will include this coupling to the $Z$ 
(although, in practice, it has a very small effect on the results presented).
It is important to point out that this is an assumption.  One could also include
an operator coupling to the $SU(2)$ field tensor, leading to a very different
value for the $Z-EDM$.  Rather than deal with two parameters, however, we just
assume that the latter operator is smaller.  If it is large, it will (barring
fine-tuning) just make the cross-section even bigger.  Note that the precise
value of the $Z-EDM$, unless it is much larger than we have assumed, has very 
little effects on the results.
The lepton-lepton-photon interaction is  $-ie\bar{\psi}
(\gamma_\mu+D\sigma_{\mu\nu}\gamma_{ 5}q^\nu)\psi A^{\mu}$, where $q$ is the 
momentum-transfer.  The electric dipole moment is defined to be $eD(q^2=0)$.
We will assume that $D(q^2)$ does not vary rapidly with $q^2$, as is the case
in virtually all models.  

The matrix element for heavy lepton production in $e^+e^-$ annihilation is given by
\begin{eqnarray}
{\cal M}&=&{ie^2\over s}\bar{v}_e\gamma^\mu u_e\bar{u}_L(
\gamma_\mu+D\sigma_{\mu\nu}\gamma_{ 5}q^\nu)v_L 
+ {ie^2\over \sin^22\theta_W(s-M^2_Z+i\Gamma M_Z)}
\bar{v}_e\gamma^\mu(C_V-C_A\gamma_{ 5})u_e\nonumber \\
&\times& \bar{u}_L\left(\gamma_\mu(C_V-C_A\gamma_{ 5} )+D\tan\theta_W\sigma_{\mu\nu}\gamma_{
5}q^\nu\right)v_L \end{eqnarray} where $C_V={1\over 2}-2\sin^2\theta_W$ and $C_A={1\over 2}$.

The differential cross section is found to be
\begin{equation}
{d\sigma\over d\Omega}={\alpha^2\over 4s}\sqrt{1-{4M^2\over s}}\left( A_1 + 
{1\over 16\sin^42\theta_W} P_{ZZ}\ A_2 +
{1\over \sin^22\theta_W} P_{\gamma Z}\ A_3\right)
\end{equation}
where
\begin{eqnarray} A_1&=&
1+\cos^2\theta + {4M^2\over s}\sin^2\theta + D^2s\sin^2\theta(1+{4M^2\over s})\nonumber \\
A_2&=&
1+\cos^2\theta-{4M^2\over s}\sin^2\theta + 4D^2s\tan^2\theta_W(\sin^2\theta
+{4M^2\over s}(1+\cos^2\theta))\nonumber \\
A_3&=&
\sqrt{1-{4M^2\over s}}\cos\theta + C_VD^2s\tan\theta_W(\sin^2\theta+
{4M^2\over s}(1+\cos^2\theta))\nonumber \\
P_{ZZ}&=&{s^2\over (s-M^2_Z)^2+\Gamma^2M^2_Z}\nonumber \\
P_{\gamma Z}&=& {s(s-M^2_Z)\over (s-M^2_Z)^2+\Gamma^2M^2_Z}
\end{eqnarray}

\begin{figure}
\begin{center}
\ \hskip -3cm \epsfysize 2in\epsfbox{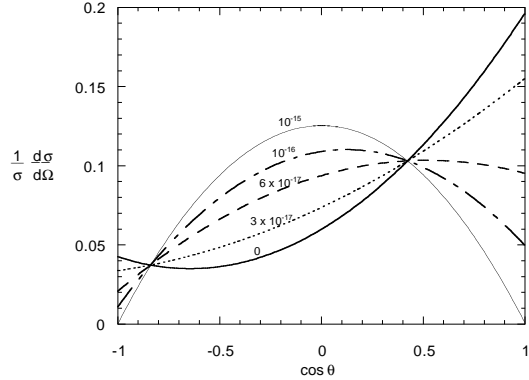}
\caption{Differential cross section for heavy charged lepton production for various EDMs, in
units of e-cm, for a lepton mass of 100 GeV.}
\end{center}
\end{figure}

In these expressions, we have dropped terms proportional to $C_V^2$ since they are
numerically negligible.   This result is plotted in Figure 1, for $M=100$ GeV and for
different values of the EDM (in units of e-cm).  Changing the lepton mass does not change
the qualitative feature of the results.  We see that for $D=0$, the usual
$1+cos^2\theta + C \cos\theta$ distribution for a lepton (where the $\cos\theta$ term is due to $\gamma-Z$ interference) is found.  For an EDM greater than 
$10^{-16}$ e-cm, the distribution is completely dominated by the $\sin^2\theta$ contribution from the electric dipole term.  (It turns out that the EDM term in the $Z$ coupling is barely noticeable in the figures.)  One can see that EDM's of the size noted earlier will dramatically alter the angular distribution.

\begin{figure}
\begin{center}
\ \hskip -2cm \epsfysize 3in\epsfbox{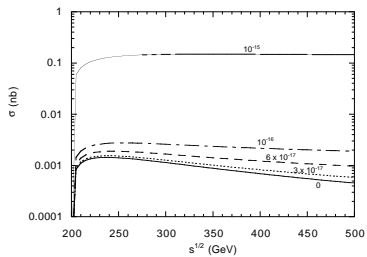}
\caption{Total cross section for heavy charged lepton production for various EDMs, with a
lepton mass of 100 GeV.}
\end{center}
\end{figure}

Of course, a scalar field will also give a $\sin^2\theta$ distribution.  Could
one distinguish this heavy lepton from, say, a heavy scalar lepton?  This can
be done easily by looking at the total cross-section, shown in Figure 2 for a
$100$ GeV lepton.   For an EDM of $10^{-16}$ e-cm, the cross section for a $\sqrt{s}=500$ GeV collider is quite large, substantially larger than expected for
a heavy fermion (and even larger than for a heavy scalar).  Note the unusual scaling
behavior.  This is not surprising.  The cross-section, for large EDM's, must
vary as $D^2$.  Thus, instead of falling as $1/s$, it becomes constant.  Thus,
by examining the threshold behavior, one could distinguish this model from
any alternatives.

Note that the cross-section varies as $D^2$, and for an EDM as large as
$1.0$ e-fermi would be almost a microbarn!!  Of course, cross sections this
large will violate unitarity.  The unitarity limit can be approximately 
estimated by setting the relevant effective interaction strength, $\alpha D\sqrt{s}$ equal to unity.  For $\sqrt{s}=500$ GeV, $D=10^{-15}$ e-cm, 
and $\alpha\simeq 1/125$, this effective interaction strength is $0.25$.  Thus
the unitarity limit will be near the upper right corner of Figure 2.
So, the behavior discussed above will appear for a wide range of parameter space without
violating the unitarity bound.  Another way of saying this is to note that 
the larger the EDM, the smaller the scale at which the physics responsible for
the effective interaction sets in, and for an EDM larger than 
$10^{-15}$ e-cm, that scale is less than $\sqrt{s}$.

These results are not original (although the discussion generally does not 
include EDMs as large as considered here).  They are given in the context of a top quark
EDM by Bernreuther et al.\cite{bsp}, and given in the context of tau-pair 
production in Ref. \cite{bno}.  This latter paper noted how one can use CP-odd 
angular correlations to search for a tau EDM, and this method has been used
by experimentalists.  However, there has not been any discussion of the 
possibility of a large EDM for heavy neutrinos, and here we see a unique 
signature.

Since there are no more than three light active neutrinos, the neutrino 
associated with the charged heavy lepton must be heavy, with a mass of at 
least $45$ GeV.  The differential cross-section is given by
\begin{equation}
{d\sigma\over d\Omega}={\alpha^2\over 4s}\sqrt{1-{4M^2\over s}}\left( A_1 + 
{1\over 8\sin^42\theta_W} P_{ZZ}\ A_2 +
{(1-4\sin^2\theta_W)\tan\theta_W\over \sin^22\theta_W} P_{\gamma Z}\ A_3\right)
\end{equation}
where 
\begin{eqnarray} A_1&=&
D^2s\sin^2\theta(1+{4M^2\over\
 s})\nonumber \\ A_2&=&
1+\cos^2\theta -{4M^2\over s}\sin^2\theta + 8C_V\cos\theta+
D^2s\tan^2\theta_W(\sin^2\theta +{4M^2\over s}(1+\cos^2\theta))\nonumber \\
A_3&=&4D^2s(\sin^2\theta+
{4M^2\over s}(1+\cos^2\theta))
\end{eqnarray}

\begin{figure}
\begin{center}
\ \hskip -2cm \epsfysize 3in\epsfbox{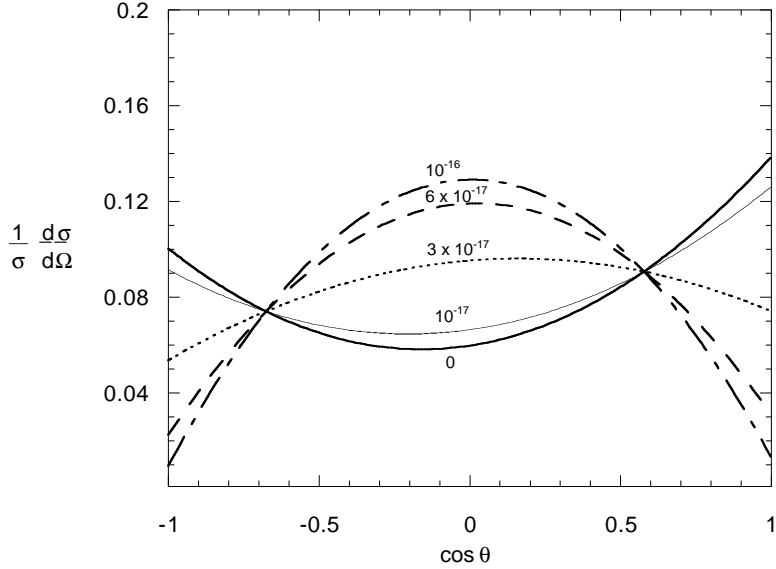}
\caption{Differential cross section for heavy neutrino production for various EDMs, in
units of e-cm, for a lepton mass of 100 GeV.}
\end{center}
\end{figure}
The differential and total cross-sections are given in Figures 3 and 4, for 
a heavy neutrino mass of $100$ GeV.  The results are similar to the charged 
heavy lepton case.  Since it is generally believed that these heavy neutrinos
could not be detected directly, such a calculation would only be meaningful
if the heavy neutrino were heavier than the charged lepton, and could thus
be detected via its decay.  Here, one could again look for CP-odd correlations,
as discussed in Ref. \cite{bno}.   However, most models have the charged
lepton heavier than the neutrino, and thus the decay can only occur through
mixing with the very light neutrinos.   As discussed in detail in Ref. \cite{fhs}, this mixing could be very small, and these heavy neutrinos could be
effectively stable. As we will see in the next section, however, 
it may be possible to detect these neutrinos directly.

\begin{figure}
\begin{center}
\ \hskip -2cm \epsfysize 3in\epsfbox{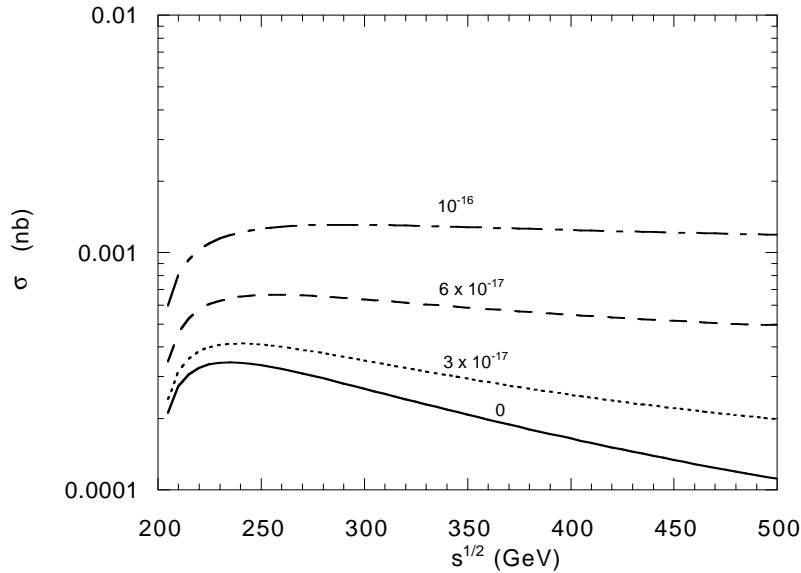}
\caption{Total cross section for heavy neutrino production for various EDMs, with the
neutrino mass of 100 GeV.}
\end{center}
\end{figure}

\section{Heavy Lepton Detection}

With one exception discussed below, the detection of heavy leptons is not substantially
affected by a large EDM.  The decays of heavy charged and neutral leptons depends
sensitively on their masses and mixings with the lighter leptons.  If the charged lepton,
$L$, has a mass $M_L>M_W+M_N$, where $M_N$ is the neutral lepton mass, then the primary
decay would be $L\rightarrow W+N$.  If the mass is less than $M_W+M_N$, but still above that
of the $N$, then its decays will either be $L\rightarrow W^*+N,W+\nu_\tau$ or $Z+\tau$,
assuming the largest mixing is with the third generation.  Which of these dominates is very
sensitive to the mixing angles (note that the $Z+\tau$ decay mode, even if rare, gives a
very clean signature).  A detailed analysis and review of all of the decays, including the
possibility that the $N$ is heavier than the $L$, is given in Ref. \cite{fhs}.  There is
little effect on their analysis from the EDM.

However, there is one exception.  It has always been assumed that heavy ($O(100)$ GeV)
neutrinos would, if stable or fairly long-lived, simply leave the detector.  But we are
interested in heavy neutrinos with a very large EDM.  Would such a neutrino leave an
ionization track in a calorimeter?  One can derive the expression for the ionization energy
loss following Jackson\cite{jackson}, replacing the electric field from a charge with the
electric field from a dipole.  Suppose a particle is traveling in the x-direction, and an
electron is at $y=b$, where $b$ is the impact parameter.  The impulse given to the electron,
$\int_{-\infty}^\infty\ E_y\ dt$ depends on the orientation of the dipole moment with the
motion of the neutrino.  If the orientation is in the $(x,y,z)$ direction, the impulse is
${e^2D\over vb^2}(0,2,2\pi)$.  Since the orientation is arbitrary, and we are only
interested in an order of magnitude estimate, we take the impulse to be ${2e^2D\over
vb^2}$.   Note that the impulse from a charge is just ${2e^2\over vb}$, and thus this result
is expected on dimensional grounds.  Converting the impulse to an energy exchange
(non-relativistically), and setting the lower bound on the impact parameter by the maximum
allowed energy exchange, one can show that the miminum impact parameter is
$b_{min}^2=e^2D/m\gamma v^2$, and thus the total energy loss is given by
\begin{equation}
{dE\over dx}= 4\pi N_A\left( {e^2\over 4\pi\epsilon_o}\right)D\gamma {Z\over A}
\end{equation}
where $N_A$ is Avogadro's number, $A$ is the atomic number in units of grams/mole.  Note
that the logarithm in the usual Bethe-Block formula is no longer present (since there is an
extra power of $b$ in the denominator of the impulse) and the electron mass and velocity
(except in $\gamma$) drop out.   Plugging in numbers, we find that ${dE\over dx}=10^{12}D\
{\rm MeV}\ {\rm g}^{-1}\ {\rm cm}^2$.  

We have seen that many models give a muon EDM of approximately $10^{-24}$ e-cm, and thus
cubic scaling would give a heavy lepton EDM of approximately $10^{-15}$ e-cm. 
In addition, models with a large dimension-five operator coefficient can have a similarly large EDM.   With that value,
the energy loss would be $10^{-3}\ {\rm MeV}\ {\rm g}^{-1}\ {\rm cm}^2$, which is roughly
one one-thousandth of the usual energy loss for a charged particle.  This would pose a
challenge, but not an insurmountable one, for experimentalists.  If the heavy neutrino is
produced in the decay of a heavy charged lepton, then it would be produced in coincidence
with a real or virtual $W$, and this would virtually eliminate backgrounds.  However, if
the heavy neutrino is produced directly through s-channel production, the backgrounds could
be formidable.  Note, however, that the cross section is huge.  For $10^{-15}$ e-cm, it is 100 picobarns, corresponding to an event rate of 1 Hertz at  the Linear Collider.   One could place very sensitive detectors some distance from the vertex, back to back, and the event rate would still be quite high.  Certainly, should a heavy charged lepton be discovered with an unusual angular
distribution, detection of a heavy neutrino though its ionization would seem possible.

\section{Conclusions}

In a number of models of CP violation,  a heavy lepton could have an enormous EDM, which could dominate the
electromagnetic properties.  The angular distribution and total cross-section for both charged
and neutral heavy leptons have been presented, and it has been noted that the heavy neutrino, if
its EDM were sufficiently large, would leave a detectable ionization track. 

In order to keep the analysis general, we have neither considered a specific model of CP
violation, and nor discussed the origin of the heavy leptons.  Although we have done the
calculation for a sequential chiral lepton doublet, the chirality of the heavy
leptons does not make much difference in our results.  In a more specific model, of course,
other issues, such as the contribution to electroweak radiative corrections, will become
relevant.

I thank Jack Kossler for several useful discussions, and Chris Carone for reading the manuscript.
This paper is dedicated to the memory of Nathan Isgur.
This work was supported by the National Science Foundation through Grant PHY-9900657.

\newpage

\def\prd#1#2#3{{\rm Phys. ~Rev. ~}{\bf D#1}, #3 (19#2)}
\def\plb#1#2#3{{\rm Phys. ~Lett. ~}{\bf B#1}, #3 (19#2) }
\def\npb#1#2#3{{\rm Nucl. ~Phys. ~}{\bf B#1}, #3 (19#2) }
\def\prl#1#2#3{{\rm Phys. ~Rev. ~Lett. ~}{\bf #1}, #3 (19#2) }

\bibliographystyle{unsrt}

\newpage

\end{document}